\documentclass[pra,twocolumn,aps,superscriptaddress]{revtex4-2}

\usepackage{graphicx}
\usepackage{dcolumn}
\usepackage{bm}
\usepackage{amssymb}
\usepackage{amsmath}
\usepackage{physics}
\usepackage{sublabel}
\usepackage[utf8]{inputenc}
\usepackage{hyperref}
\usepackage[usenames, dvipsnames]{color}
\usepackage[english]{babel}
\usepackage[T1]{fontenc}
\usepackage{bbm}
\usepackage{float}
\usepackage{verbatim}
\usepackage{tikz}
\usetikzlibrary{quantikz}
\hypersetup{colorlinks = true, linkcolor=blue, citecolor=blue, urlcolor=blue}

  \makeatletter
    \renewcommand\@make@capt@title[2]{%
     \@ifx@empty\float@link{\@firstofone}{\expandafter\href\expandafter{\float@link}}%
      {\textsc{#1}}\@caption@fignum@sep#2\quad}%
    \makeatother

\begin{document}
\title{Efficient two-qutrit gates in superconducting circuits using parametric coupling}
\author{Mahadevan Subramanian}
\email{mahadevans@uchicago.edu}
\affiliation{Department of Physics, Indian Institute of Technology Bombay, Powai, Mumbai-400076, India}

\author{Adrian Lupascu}
\email{alupascu@uwaterloo.ca}
\affiliation{Institute for Quantum Computing, Department of Physics and Astronomy, and Waterloo Institute for Nanotechnology, University of Waterloo, Waterloo, ON, Canada N2L 3G1}

\date{\today}
\begin{abstract}
    Recently, significant progress has been made in the demonstration of single qutrit and coupled qutrit gates with superconducting circuits. Coupled qutrit gates have significantly lower fidelity than single qutrit gates, owing to long implementation times. We present a protocol to implement the CZ universal gate for two qutrits based on a decomposition involving two partial state swaps and local operations. The partial state swaps can be implemented effectively using parametric coupling, which is fast and has the advantage of frequency selectivity. We perform a detailed analysis of this protocol in a system consisting of two fixed-frequency transmons coupled by a flux-tunable transmon. The application of an AC flux in the tunable transmon controls the parametric gates. This protocol has the potential to lead to fast and scalable two-qutrit gates in superconducting circuit architectures.
\end{abstract}
\maketitle

\section{Introduction} 
The framework for the theoretical exploration and applications of quantum information is usually focused on the use of two-state systems, or qubits~\cite{nielsenQuantumComputationQuantum2010}. Encoding quantum information using multilevel system, or qudits, is motivated by potential advantages in expanding the capacity of quantum information processors~\cite{wangQuditsHighDimensionalQuantum2020}, improved quantum error correction~\cite{campbellEnhancedFaultTolerantQuantum2014, gokhaleAsymptoticImprovementsQuantum2019}, and effective compilation of gates~\cite{lanyonSimplifyingQuantumLogic2009}. Besides applications in quantum computing, the use of qudits improves quantum communication~\cite{bouchardHighdimensionalQuantumCloning2017} and quantum sensing~\cite{shlyakhovQuantumMetrologyTransmon2018}, and has applications in quantum simulation~\cite{blokQuantumInformationScrambling2021, senkoRealizationQuantumIntegerSpin2015}. Currently explored physical implementations of qudits include ion traps~\cite{lowControlReadout13level2023}, molecular devices~\cite{moreno-pinedaMolecularSpinQudits2018}, solid-state defects~\cite{fuExperimentalInvestigationQuantum2022, defuentesNavigating16dimensionalHilbert2023}, and superconducting devices~\cite{blokQuantumInformationScrambling2021, yurtalanImplementationWalshHadamardGate2020a}.

Superconducting systems are a particularly favourable implementation of qudits, due to the ability to engineer quantum properties and control relevant transitions. In recent years, significant progress was made in this field, with achievements including advanced control on single~\cite{yurtalanImplementationWalshHadamardGate2020a, wuHighFidelitySoftwareDefinedQuantum2020, kononenkoCharacterizationControlSuperconducting2021, morvanQutritRandomizedBenchmarking2021} and coupled~\cite{https://doi.org/10.48550/arxiv.2206.07216, https://doi.org/10.48550/arxiv.2206.11199, PhysRevApplied.19.064024} qutrits. In a manner similar to qubit based computing, two qudit gates have significantly larger errors than single qutrit gates \cite{PhysRevX.13.021028}, owing to the inherently longer execution time in currently used approaches. We explore a method for qutrit-qutrit gates focused on the implementation of the universal CZ gate using an effective decomposition into swap-type gates based on parametric coupling. Parametric coupling has been used extensively for coupled qubit gates and has been applied in recent works to qutrit-qutrit gates \cite{PhysRevApplied.19.064024,PhysRevA.108.032615}. We identify optimal decompositions of a CZ gate into parametric gates, and we performed a detailed analysis for transmon based qutrits.

The contents of this paper is divided into three sections. In the following section, we develop and discuss the theory and explain the working principle of swap type gates using parametric coupling. In the next section, we perform a numerical analysis of parametric gates based on simulations of the dynamics while also discussing the challenges involved in choosing appropriate parameters for these simulations. Finally, in the next section we show how qutrit gates can be compiled using the two entangling gates we implement using parametric coupling including a way to decompose the qutrit CZ gate.

\section{Implementation of SWAP-type gates using parametric coupling}
Parametric gates are enabled by modulating the couplings or energy levels of a circuit at a specific frequency so as to enable a sideband transition between certain energy levels \cite{PhysRevApplied.6.064007,PhysRevB.87.220505,PhysRevB.73.064512,doi:10.1126/sciadv.aao3603,PhysRevApplied.16.024050,PhysRevApplied.10.034050,PhysRevA.97.022330,PhysRevB.73.094506,PhysRevA.96.062323,PhysRevA.102.022619}. 
Parametric coupling shows great promise in designing scalable superconducting circuits  \cite{PhysRevApplied.6.064007,doi:10.1126/sciadv.aao3603} by allowing desired specific transitions to be activated based on frequency selectivity. A commonly explored setup involves transmons coupled through a tunable transmon \cite{PhysRevApplied.6.064007}. Alternate circuit configurations which use transmons as well \cite{PhysRevA.97.022330,PhysRevApplied.19.064043,PhysRevApplied.18.034038} and flux qubits or DC-SQUIDs instead of transmons \cite{PhysRevB.73.064512,PhysRevB.73.094506} have been explored as well. Our theoretical analysis of parametric coupling, while focused on two transmons coupled via a flux tunable transmon, can be straightforwardly extended to alternative superconducting circuits.
\begin{figure}[ht]
    \centering
    \includegraphics[width=\linewidth]{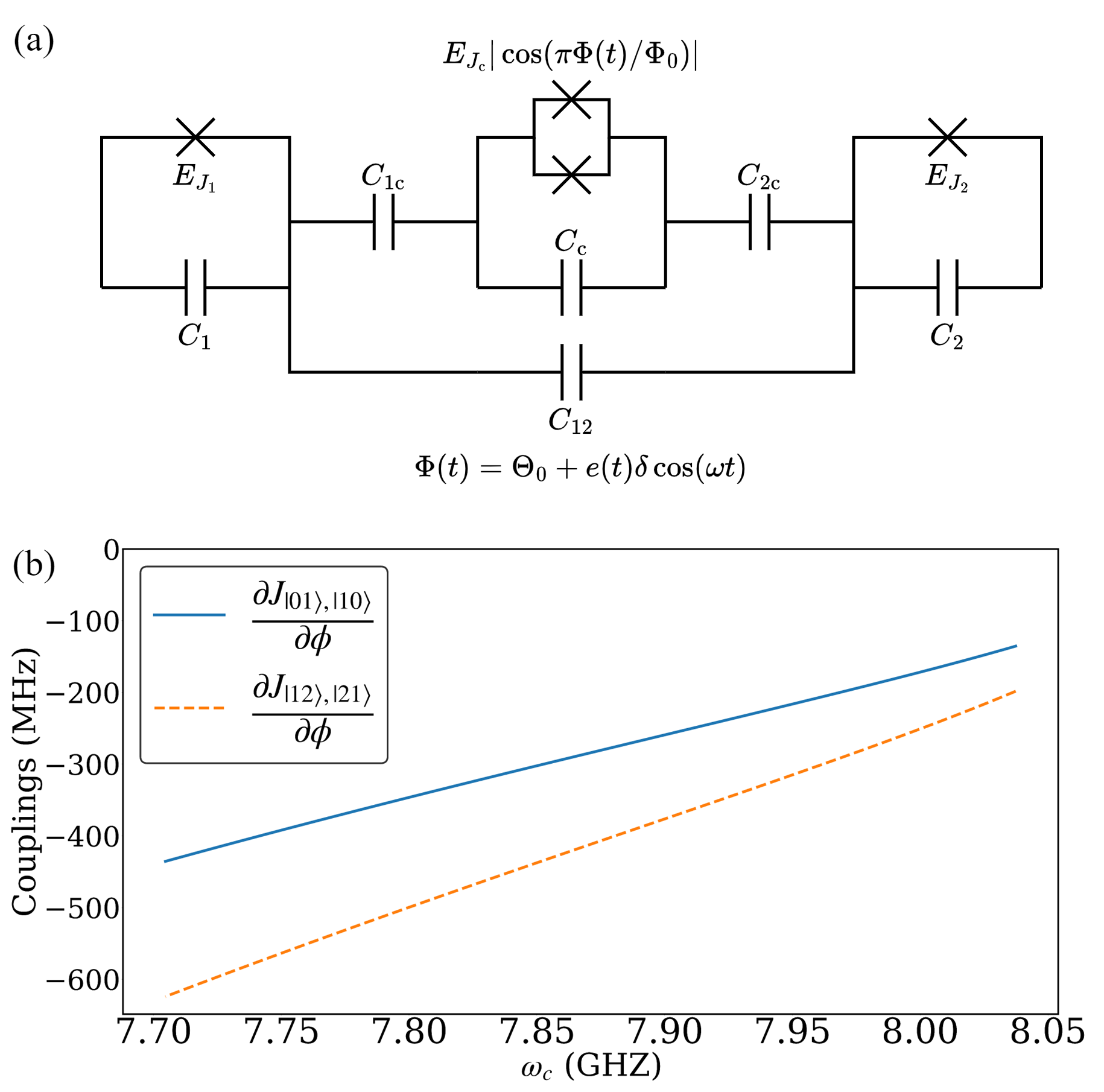}
    \caption{(a) Schematic depicting the circuit diagram of a superconducting circuit formed of two fixed-frequency transmons $\mathrm{Q1}$ and $\mathrm{Q2}$ and a tunable coupler transmon $\mathrm{C}$, coupled via capacitors. (b) The dependence of the flux derivative of the coupling between levels $\ket{01}, \ket{10}$ ($J_{\ket{01}, \ket{10}}$) and $\ket{12}, \ket{21}$ ($J_{\ket{12}, \ket{21}}$) on the coupler frequency $\omega_c$.}
    \label{fig:1}
\end{figure}
\subsection{Circuit Hamiltonian} 
The circuit that we analyze (see figure \ref{fig:1}) consists of two fixed frequency transmons ``$\mathrm{Q1}$'' and ``$\mathrm{Q2}$'' and a coupler implemented as a flux - tunable transmon ``$\mathrm{C}$''. This architecture is directly based on the use of a tunable bus \cite{PhysRevApplied.6.064007} and similar architectures have been used for two-qubit gates \cite{PhysRevApplied.10.054062, PhysRevLett.125.240503}. All the transmons are capacitively coupled to each other which results in the coupled system Hamiltonian

\begin{gather} \label{eq:LabFrameHamiltonian}
    \hat{H}_\mathrm{lab}(t) = \hat{H}_0(t) + \hat{H}_{m} + \hat{H}_{d}
\end{gather}
with 
\begin{gather}
 \begin{aligned}
        \hat{H}_0 = \sum_{i=1,2,\mathrm{c}}\left(E_{C_i}\hat{n}_i^2 - E_{J_i}\cos{\hat{\phi}_i}\right)
    \end{aligned}, \\
    \hat{H}_{\mathrm{m}} = g_1\hat{n}_1\hat{n}_{\mathrm{c}} +g_2\hat{n}_2\hat{n}_{\mathrm{c}} ,
\end{gather}
and
\begin{gather}
    \hat{H}_{\mathrm{d}} = g_{12}\hat{n}_1\hat{n}_2 .
\end{gather}

\indent Here we consider the tunable transmon $\mathrm{C}$ junction to have zero asymmetry and so the Josephson energy $E_{J_{\mathrm{c}}} = E_{J_{\mathrm{c}_0}}|\cos(\Phi(t))|$. The constants $g_1$, $g_2$ and $g_{12}$ are dependent on the capacitance values and are explicitly calculated in \cite{PhysRevApplied.10.054062} to be:
\begin{equation}
    \begin{aligned}
        g_{i} &= 8\frac{C_{ic}}{\sqrt{C_iC_c}}\sqrt{E_{C_i}E_{C_C}} \text{, and}\\
        g_{12} &= 8\left(1 + \frac{C_{1c}C_{2c}}{C_{12}C_{c}}\right)\frac{C_{12}}{\sqrt{C_1C_2}}\sqrt{E_{C_1}E_{C_2}}.
    \end{aligned}
\end{equation}
The values of these parameters are chosen such that $g_{12}\ll g_1,g_2$ and $g_1, g_2 \ll \Delta$. Here $\Delta$ is a scaling parameter which is of the same order as the difference between the energies of the first excited state and of the ground state  of $\hat{H}_0$. 
The flux being a function of time is taken to be of the form
\begin{equation}\label{eq:fluxT}
    \Phi(t) = \Theta_0 + e(t)\delta\cos(\omega t),
\end{equation}
where  $\Theta_0$ is an offset and $e(t)$ is an envelope function that is nonvanishing for times $0 \leq t \leq T$ and fulfills $e(t=0) = e(t=T) = 0$. This flux pulse offers an AC control to the circuit. The value of $\omega$ can be chosen such that  transitions are induced between certain levels of the three-body system formed of $\mathrm{Q1}$, $\mathrm{Q2}$ and $\mathrm{C}$. The values of $\delta$ and $\Theta_0$ adjust the value of the effective coupling for this transition as discussed in detail in the following subsections.

\subsection{Effective Hamiltonian} 
In this subsection we discuss the derivation of an effective Hamiltonian for the system. Obtaining the effective Hamiltonian requires performing an appropriate frame transformation involving the time-dependent Schrieffer-Wolf transformation \cite{PhysRevA.96.062323, PhysRevA.95.022314}. The Schrieffer-Wolff transformation \cite{BRAVYI20112793} leads to a new frame, where the Hamiltonian is block diagonal with respect to certain subsystems to a good approximation. Here coupling terms involving the transmon $\mathrm{C}$ are eliminated, hence capturing the effective dynamics of $\mathrm{Q1}$ and $\mathrm{Q2}$.\\
\indent We use the frame transformation $\exp(S(t))$, where $S(t)$ is an anti-Hermitian operator. Applying this transformation to the Hamiltonian in Eq.~\ref{eq:LabFrameHamiltonian} leads to a Hamiltonian in the new frame given by
\begin{equation}
    \begin{aligned}
        \hat{H}_\mathrm{SW}(t) &= e^{S(t)}(\hat{H}_\mathrm{lab}-i\partial_t)e^{-S(t)}\\ &= e^{S(t)}(\hat{H}_\mathrm{lab}+i(\partial_tS(t)))e^{-S(t)} .
    \end{aligned}
\end{equation}
\indent This frame will be referred to as the SWT frame. The operator $S(t)$ must result in an appropriate transformation such that at any time $\hat{H}_{\mathrm{SW}}(t)$ is block diagonal over $\mathrm{Q1}$ and $\mathrm{Q2}$ without any interactions with $\mathrm{C}$. Exact analytical solutions for $S(t)$ can be obtained in some cases by the use of an ansatz based approach \cite{https://doi.org/10.48550/arxiv.2004.06534}. A more general approach is to expand $S(t)$ as a series $S = \sum_{i=1}^{\infty}S^{(i)}$. This series is such that $S^{(i)}$ scales as $\lambda^i$ where $\lambda$ is a small parameter, characteristic of the coupling. In our case, we can expand the Hamiltonian as three terms $\hat{H}_{\mathrm{lab}} = \hat{H}_0 + \hat{H}_{\mathrm{m}} + \hat{H}_{\mathrm{d}}$. We wish to eliminate all terms coming from $\hat{H}_{\mathrm{m}}$ and since we already have $g_1,g_2 \ll \omega_0$, we define the order parameter to be $\lambda = g_1/\omega_0$ (it is assumed that $g_1\sim g_2$). Using the Baker–Campbell Hausdorff (BCH) expansion we obtain
\begin{align}
    \hat{H}_{\mathrm{SW}}(t) &= \sum_{i=0}^\infty\hat{H}_{\mathrm{SW}}^{(i)}(t),
    \intertext{with}
    \hat{H}_{\mathrm{SW}}^{(0)}(t) &= \hat{H}_0(t),\\
    \hat{H}_{\mathrm{SW}}^{(1)}(t) &= \hat{H}_{\mathrm{m}} + [S^{(1)}(t), \hat{H}_0(t)] + i\dot{S}^{(1)}(t)
\end{align}
and
\begin{equation}
    \begin{aligned}
        \hat{H}_{\mathrm{SW}}^{(2)}(t) &= \hat{H}_{\mathrm{d}} + \frac{1}{2}[S^{(1)}(t), [S^{(1)}(t), \hat{H}_0(t)]] + [S^{(1)}(t), \hat{H}_{\mathrm{m}}] \\&+ [S^{(1)}(t),i\dot{S}^{(1)}(t)] + [S^{(2)}(t), \hat{H}_0(t)] + i\dot{S}^{(2)}(t).
    \end{aligned}
\end{equation}
\indent By imposing the condition that each of the terms $\hat{H}_{\mathrm{SW}}^{(i)}(t)$ has zero coupling terms involving transmon $\mathrm{C}$, we obtain differential equations for each of the time-dependent $S^{(i)}$ terms. For simplification we represent the Hamiltonian as $\hat{H}_\mathrm{SW}(t) = \hat{H}_\mathrm{SW,0}(t) + \hat{H}_\mathrm{SW,c}(t)$. Here $\hat{H}_{\mathrm{SW,0}}(t)$ represents the diagonal part of $\hat{H}_{\mathrm{SW}}(t)$ and $\hat{H}_\mathrm{SW,c}(t)$ represents the off-diagonal part when expressed in the basis which diagonalizes the lab Hamiltonian not including coupling terms, which is given by $\hat{H}_0(t=0)$.\\

\indent To further simplify the analysis of the dynamics, we introduce a new rotating frame. This  rotating frame is defined, relative to the lab frame, using the transformation $\ket{\psi_\mathrm{rot}(t)} = U_\mathrm{rot}(t)\ket{\psi_\mathrm{lab}(t)}$, where
\begin{equation}\label{eq:finalTransform}
    U_\mathrm{rot}(t) = \exp\left(\frac{i}{\hbar}\int_0^t \hat{H}_\mathrm{SW,0}(t')dt'\right)\exp(S(t)).
\end{equation}
This transformation essentially takes the state to the SWT frame and then rotates it so as to eliminate the diagonal terms in the Hamiltonian. The unitary operation in in Eq.~\eqref{eq:finalTransform} is a transformation from the lab frame to the final frame, which is used for computation. The Hamiltonian in this final frame is given by
\begin{equation}
    \begin{aligned}
    \hat{H}_\mathrm{rot}(t) &= U_\mathrm{rot}(t)H_\mathrm{lab}(t)U^\dagger_\mathrm{rot}(t) - i\hbar U_\mathrm{rot}(t)\partial_tU_\mathrm{rot}^\dagger(t)\\
    &= e^{i\mathcal{I}(t)}\hat{H}_\mathrm{SW,c}(t)e^{-i\mathcal{I}(t)},
    \end{aligned}
\end{equation}
where $\mathcal{I}(t) = \int_0^t \hat{H}_\mathrm{SW,0}(t')dt'/\hbar$.

\subsection{Parametrically activated entangled gates}
\begin{figure}[ht]
    \centering
    \includegraphics[width=\linewidth]{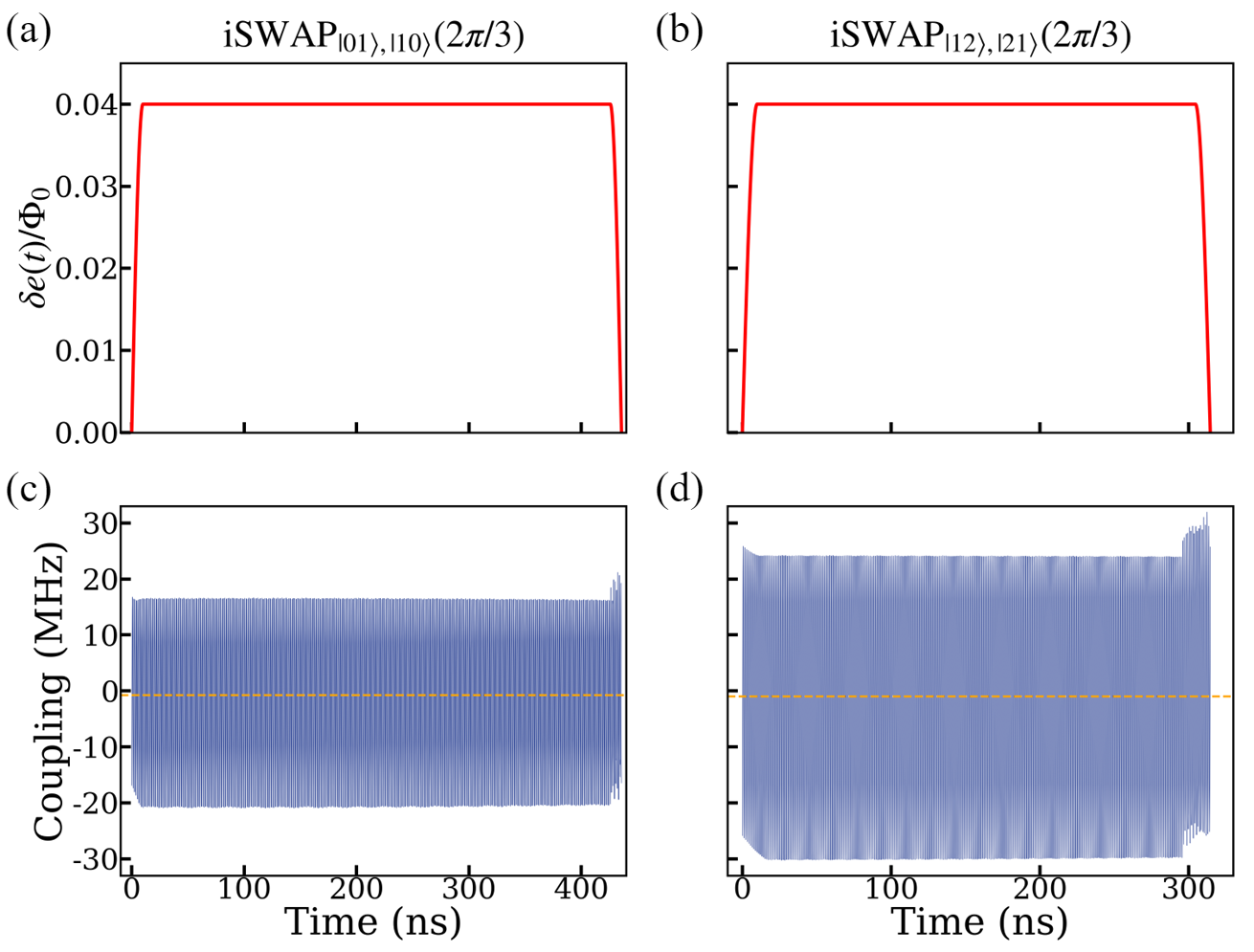}
    \caption{(a) Envelope of the pulse used for the iSWAP$_{01,10}(2\pi/3)$ gate. (b) Envelope of the pulse used for the iSWAP$_{12,21}(2\pi/3)$ gate. (c) The value of the coupling between levels $\ket{01}$ and $\ket{10}$ during the application of the control pulse for iSWAP$_{01,10}(2\pi/3)$ which shows slight displacement from a mean point of zero depicting the value of effective coupling (mean value shown as the dashed orange line). (c) The value of the coupling between levels $\ket{12}$ and $\ket{21}$ during the application of the control pulse for iSWAP$_{12,21}(2\pi/3)$, with the mean value as dashed orange line.}
    \label{fig:2}
\end{figure}

We now study the dynamics when applying a time dependent flux, as described in equation \eqref{eq:fluxT}. The envelope we use has sinusoidal rising and falling shaped as seen in figures \ref{fig:2}(a) and \ref{fig:2}(b). We define $e(t)$ as follows
\begin{equation}
    e(t) = \begin{cases}\sin\left(\frac{\pi t}{2T_{\mathrm{rise}}}\right) & 0\leq t\leq T_{\mathrm{rise}}\\
    1 & T_{\mathrm{rise}}\leq t\leq T - T_{\mathrm{fall}}\\
    \sin\left(\frac{\pi (T-t)}{2T_{\mathrm{fall}}}\right) & T - T_{\mathrm{fall}\leq t\leq T}\end{cases}
\end{equation}
For sake of simplification, we first examine the case where the rise time $T_{\mathrm{rise}}$ and fall time $T_{\mathrm{fall}}$ are zero. Given that $\hat{H}_{\mathrm{SW,c}}(\Phi)$ is flux dependent, we can perform a Taylor expansion, which leads to the following Hamiltonian:
\begin{equation}
    \begin{aligned}
    \hat{H}_{\mathrm{SW,c}}(\Phi) &= \hat{H}_{\mathrm{SW,c}}(\Phi_0) + \frac{\delta^2}{4}\frac{\partial^2\hat{H}_{\mathrm{SW,c}}(\Phi_0)}{\partial\Phi^2} \\&\quad+ \delta\cos{\omega t}\left(\frac{\partial\hat{H}_{\mathrm{SW,c}}(\Phi_0)}{\partial\Phi}\right) \\&\quad+ \frac{\delta^2\cos{(2\omega t)}}{4}\left(\frac{\partial^2\hat{H}_{\mathrm{SW,c}}(\Phi_0)}{\partial\Phi^2}\right) + \mathcal{O}(\delta^3)
\end{aligned}
\end{equation}
\indent We choose our computational basis to be the basis set $\{\ket{i}\}$ which diagonalizes $\hat{H}_0(\Phi_0)$. The matrix elements of $\hat{H}_\mathrm{rot}(t)$ can be expressed as
\begin{equation}
    \langle i|\hat{H}_\mathrm{rot}(t)|j\rangle = \sum_{k,l}\langle i|e^{i\mathcal{I}(t)}|k\rangle\langle k|\hat{H}_\mathrm{SW,c}(t)|l\rangle\langle l|e^{-i\mathcal{I}(t)}|j\rangle.
\end{equation}
Trivially, $\mathcal{I}(t)$ is diagonal in this chosen basis and so we define 
\begin{equation}
\mathcal{I}(t)_{i,i} = \langle i|\mathcal{I}(t)|i\rangle.\end{equation} We can approximate $\mathcal{I}(t)_{i,i} \approx \omega_i t$ where $\omega_i$ is the eigenvalue of $\hat{H}_0(\Phi_0)$ for the eigenvector $\ket{i}$ with a small added correction of $\mathcal{O}(\delta^3)$ (see appendix \ref{ap:Eqrotframe}). This is due to the  integral of $\hat{H}_{\mathrm{SW,0}}$ being dominated by the constant terms over large enough time frames and additionally by the terms which bring a time-dependence having smaller magnitudes when the coupler transmon is far detuned from the transmons Q1 and Q2. A detailed discussion on this point can be found in appendices \ref{ap:rotframe} and \ref{ap:Eqrotframe}. This leads to the simplification of the expression of the matrix element as $\langle i|\hat{H}_\mathrm{rot}(t)|j\rangle = \langle i|\hat{H}_\mathrm{SW,c}(t)|j\rangle e^{i(\omega_i-\omega_j)t}$. After elimination of the rotating terms, the condition $\omega = \omega_i-\omega_j$ gives us the following equivalent Hamiltonian: 
\begin{equation}
    \hat{H}_{\mathrm{equiv}} = \left(\frac{\partial\langle i|\hat{H}_{\mathrm{SW,c}}(\Phi_0)|j\rangle}{\partial\Phi}\right)\frac{(\ket{i}\bra{j} + \ket{j}\bra{i})}{2} .
\end{equation}
It must be noted that this equivalent form of the Hamiltonian works under an assumption that $\Phi(t) = \Theta_0 + \delta\cos(\omega t)$ from $t=0$ to $t=T$. If we instead assume $\Phi(t) = \Theta_0 + \delta\cos(\omega t + \phi)$, where $\phi$ is the phase of the driving, a factor of $e^{i\phi}$ arises in the equivalent Hamiltonian as can be seen in the following equation.
\begin{equation}\label{eq:finalequiv}
    \hat{H}_{\mathrm{equiv},\phi} = \left(\frac{\partial\langle i|\hat{H}_{\mathrm{SW,c}}(\Phi_0)|j\rangle}{\partial\Phi}\right)\frac{(e^{-i\phi}\ket{i}\bra{j} + e^{i\phi}\ket{j}\bra{i})}{2} .
\end{equation}
Due to the rise and fall times, even with a pulse which is having $\phi=0$, the period where the envelope is held at $e(t)=1$ can be isolated to be an interaction with some non-zero $\phi$ causing the final result to be having a certain phase. This effect can be fixed in experiment by choosing an appropriate $\phi$ to effectively cancel out this effect or by making use of the appropriate single qutrit virtual Z \cite{PhysRevA.96.022330} which can also directly correct this.\\
\indent We introduce two families of gates, denoted by $\mathrm{iSWAP}_{01,10}(\theta)$ and $\mathrm{iSWAP}_{12,21}(\theta)$. These gates are the gates obtained by a $\theta$ rotation when using the Hamiltonian $\hat{H}_{\mathrm{equiv}}$ with $\ket{i},\ket{j}$ being $\ket{01},\ket{10}$ and $\ket{12},\ket{21}$ respectively. These gates are iSWAP gates over the two states when $\theta=\pi$. Shown in figure \ref{fig:2}(c) and \ref{fig:2}(d) are the effective coupling terms between levels $\ket{01},\ket{10}$ and $\ket{12},\ket{21}$ respectively. These are oscillating around some fixed mean value and while the oscillations will effectively cancel out, the mean value is non-vanishing and it is due to the first order term in the Taylor expansion becoming non-rotating due to modulation at the correct frequency. This is what results in an effective coupling between the two levels.

\section{Numerical studies of parametric gates}

In this section we analyze the qutrit-qutrit parametric gates introduced in the previous section using numerical simulations of the time dynamics for a realistic superconducting circuit. The analysis considers ways to avoid the potential spurious effects related to frequency crowding and leakage to higher levels. Frequency crowding can greatly reduce fidelity in parametric gates since it may result in unwanted interactions being activated along with the intended transition.\\
\indent For the $\ket{01}$ to $\ket{10}$ exchange, the difference between the levels is given approximately by $\Delta_{01,10}\approx\omega_2 - \omega_1$ where $\omega_1$ and $\omega_2$ are the frequencies of the transition from ground state to the first excited state of the transmons $\mathrm{Q1}$ and $\mathrm{Q2}$ respectively. The levels $\ket{12}$ and $\ket{21}$ have a frequency difference $\Delta_{12,21} \approx \omega_1-\omega_2 + \alpha_1-\alpha_2$. If $\alpha_1-\alpha_2$ is not large enough, choosing $\omega=\Delta_{01,10}$ can result in slightly activating the exchange between $\ket{12}$ and $\ket{21}$. Similarly an appropriate anharmonicity must be chosen to avoid exchanges such as those between $\ket{11}$ and $\ket{02}$ or $\ket{20}$ when using $\omega=\Delta_{01,10}$. Our analysis focuses on choosing an appropriate parameter set which is realistic enough to be experimentally verified while still offering as much high fidelity as possible.

\begin{figure}[ht]
    \centering
    \includegraphics[width=\linewidth]{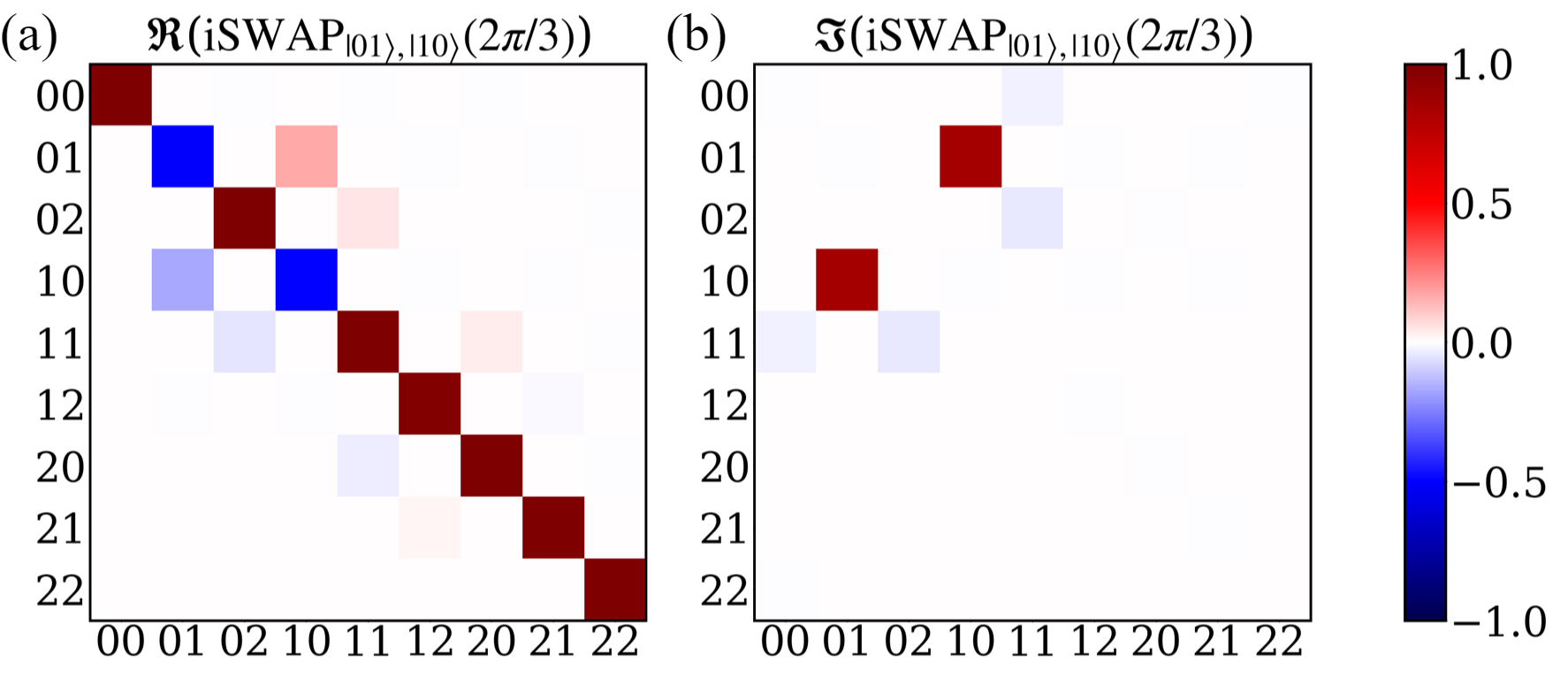}
    \caption{$\mathrm{iSWAP}_{01,10}(2\pi/3)$ with 99.65\% fidelity with length of 315 ns including 10 ns of both rise and fall time. Shown plotted above are the (a) real part of the unitary and (b) imaginary part of the unitary. In both (a) and (b), the matrix element shown is the $\langle y|U|x\rangle$ where $x$ and $y$ are the states corresponding to the respective label on the $x$ and $y$ axes. These graphs visually depict how the implemented unitary deviates from the ideal mainly with the by all the non diagonal matrix elements which are not between the 01 and 10 states. In this particular case one can clearly see the fidelity is decreased due to exchange between 11-02, 11-20, and 12-21.}
    \label{fig:3}
\end{figure}

\begin{figure}[ht]
    \centering
    \includegraphics[width=\linewidth]{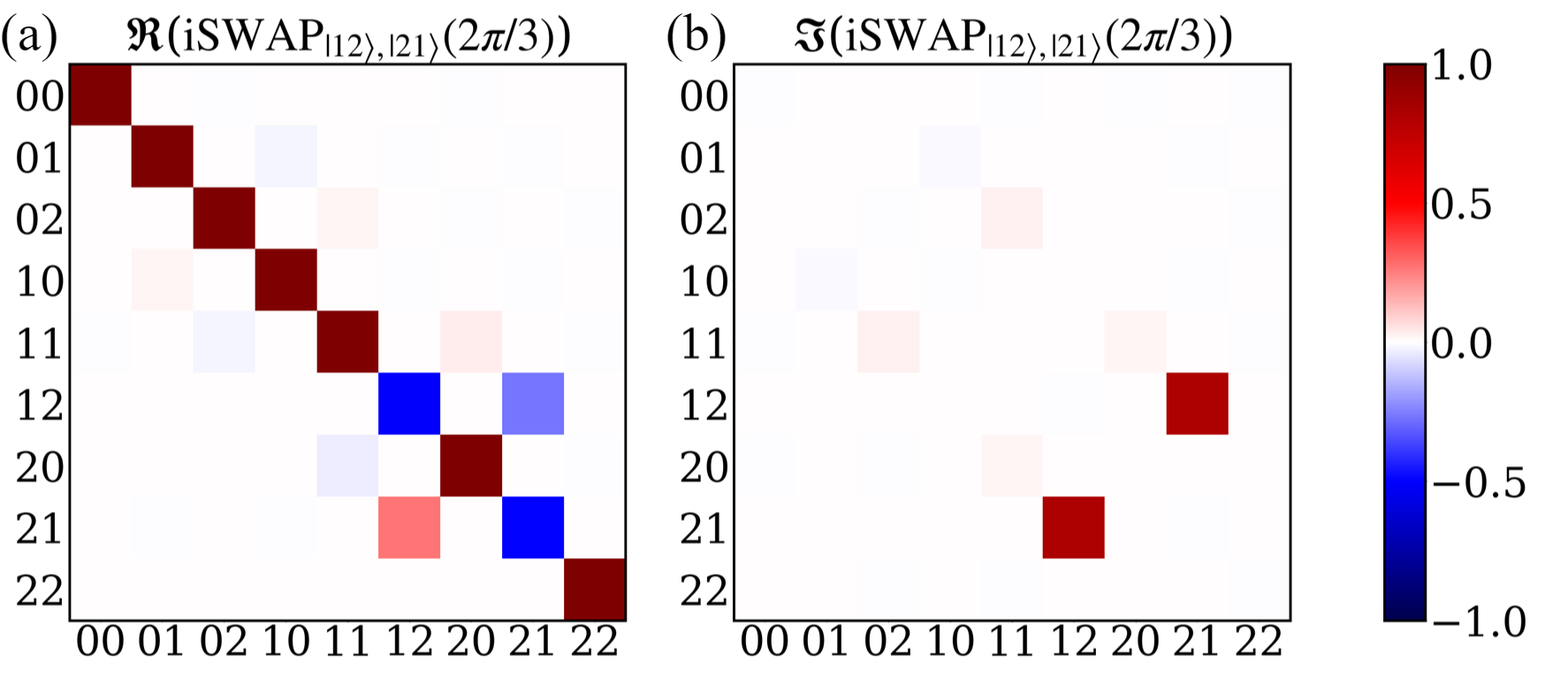}
    \caption{$\mathrm{iSWAP}_{12,21}(2\pi/3)$ with 99.72\% fidelity with length of 435 ns including 10 ns of both rise and fall time. Shown plotted above are the (a) real part of the unitary and (b) imaginary part of the unitary. In both (a) and (b), the matrix element shown is the $\langle y|U|x\rangle$ where $x$ and $y$ are the states corresponding to the respective label on the $x$ and $y$ axes. Again we can see how the implemented unitary deviates from the ideal mainly with the by all the non diagonal matrix elements which are not between the 12 and 21 states. Here the fidelity can be seen to have been decreased due to exchange between 11-02, 11-20, and 10-01}
    \label{fig:4}
\end{figure}

Based on our analysis keeping in mind the above mentioned constraints, we select the following parameters for the Hamiltonian described in equations (2) and (3).
\begin{table}[ht]
    \centering
    \begin{tabular}{|c|c|c|c|c|c|c|c|c|c|c|}
    \hline
      Parameters & $E_{J_1}$ & $E_{C_1}$ & $E_{J_2}$ & $E_{C_2}$ & $E_{J_{\mathrm{c}}}$ & $E_{C_{\mathrm{c}}}$ & $g_1$ & $g_2$ & $g_{12}$\\
       \hline
       Value (GHz) & 13.5 & 0.28 & 20.5 & 0.24 & 39.5 & 0.220 & 0.146 & 0.164 & 0.015\\
    \hline
    \end{tabular}
    \caption{Simulation parameters}
    \label{tab:1}
\end{table}
The simulations are carried out using the QuTip \cite{JOHANSSON20121760,JOHANSSON20131234} and scqubits \cite{Groszkowski2021scqubitspython} packages.\\
\indent Based on the final equivalent Hamiltonian we obtain in \eqref{eq:finalequiv}, we  expect that the rotation angle of a general SWAP operation is directly proportional to the duration of the pulse. However, deviations from a linear dependence are observed which are related to the  finite rising and falling times of the pulses and to the phase of the pulse at each instance resulting in a slightly non-linear dependence. To implement SWAP rotations with precise angles, a rotation time is calculated assuming that the linear dependence holds and then a search in a range around this value is used to find pulse times that correspond to the desired rotation angles. The obtained unitary gates are depicted in figure \ref{fig:3} for a $\mathrm{iSWAP}_{01,10}(2\pi/3)$ with 99.65\% fidelity and in figure \ref{fig:4} for a $\mathrm{iSWAP}_{12,21}(2\pi/3)$ with 99.72\% fidelity.

\section{Compilation of qutrit gates}
\indent Universal control in qudits is achieved by using a gate set that includes arbitrary single qudit rotations and one suitable entangling two-qudit gate \cite{brylinski2002mathematics}. If there exists a product state in $(\mathbb{C^d})^{\otimes 2}$ which is mapped to an entangled state in $(\mathbb{C^d})^{\otimes 2}$ by the use of a unitary $V$, that unitary is called imprimitive or entangling \cite{brylinski2002mathematics}. Crucially, the conditions for universality highlighted in ref.~\cite{brylinski2002mathematics} require only one kind of entangling gate but these do not place restrictions on the number of gates. This is where we employ two kinds of entangling gates, namely the rotations iSWAP$_{01,10}$ and iSWAP$_{12,21}$ which can cover a greater variety of two-qutrit gates with lower circuit depths due to the additional degree of freedom.\\
\indent To understand the applicability of the gates we obtain in this protocol, we apply them to the task of compiling two-qutrit gates. We decompose a qutrit CZ gate defined as
\begin{equation}
    \mathrm{CZ} = \sum_{j,k\in\{0,1,2\}}(\exp(2i\pi/3))^{jk}\ket{jk}\bra{jk},
\end{equation}
which is universal, into single qutrit rotations and rotations of iSWAP$_{01,10}$ and iSWAP$_{12,21}$.
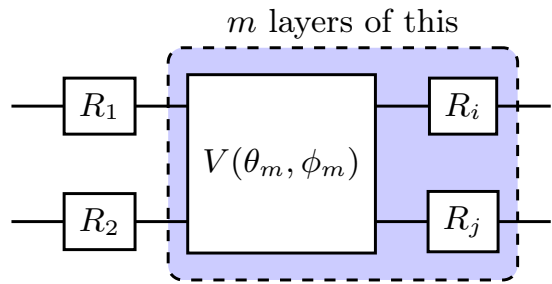
\begin{figure}
    \centering
\scalebox{1.4}{
\begin{quantikz}
& \gate{R_1} & \gate[2,nwires={1,2}]{V(\theta_m,\phi_m)}\qw\gategroup[wires=2,steps
=2,style={dashed,
rounded corners,fill=blue!20, inner xsep=2pt},background]{$m$ layers of this} & \gate{R_i} & \qw \\
& \gate{R_2} & \qw & \gate{R_j} & \qw
\end{quantikz}}
\caption{Circuit ansatz for gate compilation. We use a known 2 qudit gate $V$ which takes two angles as input and maximize the fidelity $\mathcal{F}(U_{targ},U)$ where $U$ is the unitary of the circuit. The operators $R_{i}$ are arbitrary $SU(3)$ rotations which have the variables to be optimized. For qutrits, $m$ layers have optimization is over $18m$ variables.}
\label{fig:ansatz}
\end{figure}
 Making use of the parameterized quantum circuit which is described in figure \ref{fig:ansatz}, we perform an optimization over fidelity of achieved unitary and target unitary to find an optimal decomposition of the target unitary.\\
\begin{figure}[ht]
    \centering
    \includegraphics[width=\linewidth]{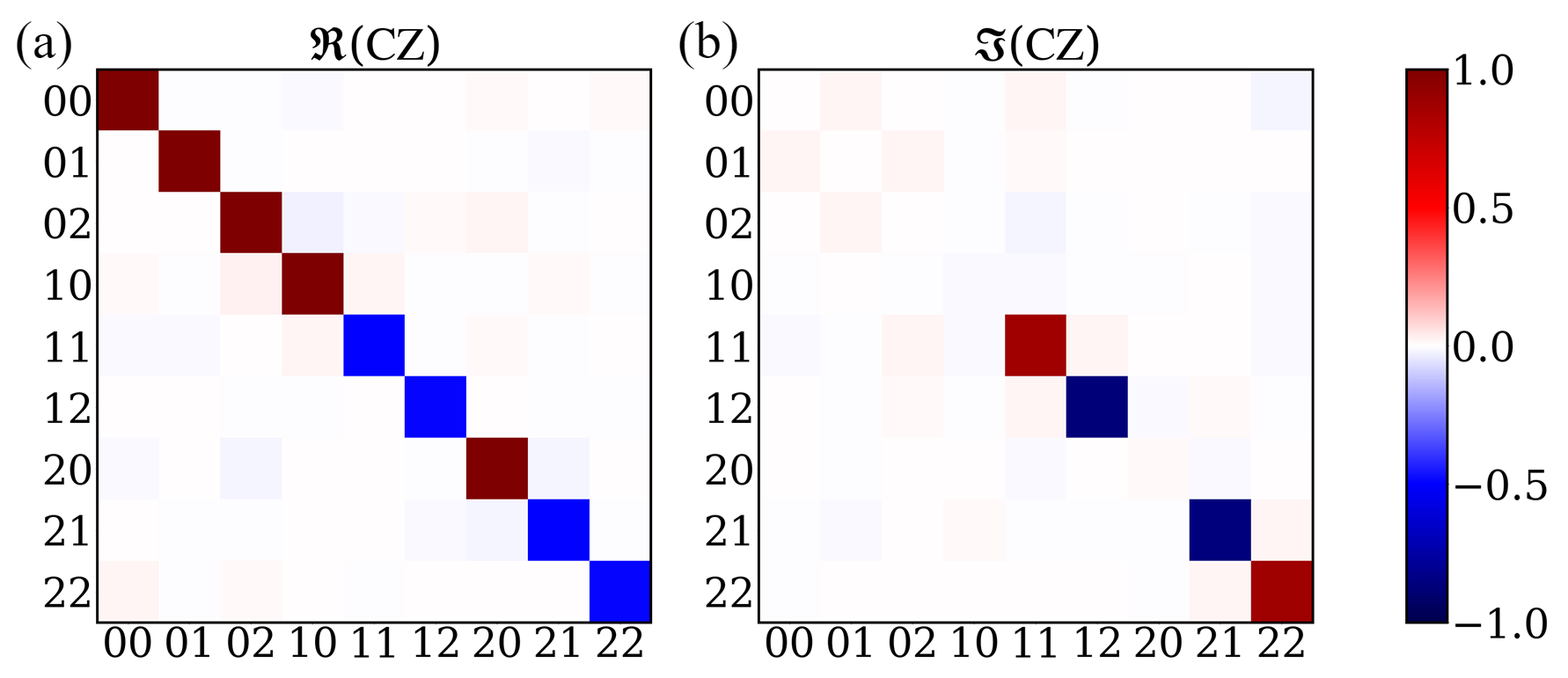}
    \caption{CZ gate with 99.41\% fidelity shown plotted above as (a) real part of the unitary and (b) imaginary part of the unitary. In both (a) and (b), the matrix element shown is the $\langle y|U|x\rangle$ where $x$ and $y$ are the states corresponding to the respective label on the $x$ and $y$ axes. The non diagonal terms give a visible indication of the imperfections of this gate which is supposed to be diagonal.}
    \label{fig:5}
\end{figure}
\indent The method uses a layered parameterized quantum circuit. The cost function is defined as 
\begin{equation}
    \mathcal{C}(\pmb{\theta}) = 1 - \frac{1}{d^2}|\mathrm{Tr}(U_{\mathrm{targ}}^\dagger U(\pmb{\theta}))|,
\end{equation} where the implemented unitary is $U(\pmb{\theta}))$, with $\pmb{\theta}$ is the set of controllable parameters in the circuit, and $d$ is the qudit dimension. It is clear that $\mathcal{C}(\pmb{\theta}) \geq 0$ for all $\pmb{\theta}$ and it equals zero iff $U(\pmb{\theta}) = U_{\mathrm{targ}}^\dagger$ (upto a global phase).\\
\indent Using two layers of such an ansatz we obtain an exact decomposition for the qutrit CZ gate. The entangling gates this uses are an $\mathrm{iSWAP}_{01,10}(-2\pi/3)$ and $\mathrm{iSWAP}_{12,21}(2\pi/3)$ in combination with single qutrit rotations. The exact values used in this decomposition for the single qutrit rotations cam be found in the supplemental material.

Using the gates obtained from simulations as shown in figures \ref{fig:3} and \ref{fig:4}, we obtain the CZ gate shown in figure \ref{fig:5} which has 99.41\% fidelity. This fidelity is calculated by assuming a perfect fidelity for single qutrit rotations and then proceed to substitute the ideal entangling gate of the CZ decomposition with an implemented version of it. In practice there could be decreased fidelity due to non-ideal single qutrit rotations. This obtained CZ gate stands as a proof of concept for how high of a fidelity can be obtained using our parameter set in simulation.
\section{Conclusion}
\indent This work explored an efficient method for implementing two-qutrit gates using transmons making use of parametric coupling. Making use of a tunable transmon as a coupler, we create entangling gates between two other fixed frequency transmons. An important result we present is the exact decomposition of the qutrit CZ gate using two specific kinds of entangling rotations. Using the method of parametric coupling, we show how we can implement a qubit iSWAP gates involving a $\ket{01}$, $\ket{10}$ exchange \cite{PhysRevApplied.6.064007} as well as an extension of the same method to obtain a similar gate for exchange between $\ket{12}$ and $\ket{21}$.\\
\indent The exact decomposition of the qutrit CZ gate in terms of the exchange interactions between $\ket{01}$ to $\ket{10}$ and $\ket{12}$ to $\ket{21}$ is obtained using a simple gate compilation technique yielding an exact solution by having conducted an optimization over rotation parameters for certain gates. Using this decomposition in conjunction with the simulated gates we show a potential CZ gate which can be obtained assuming perfect fidelity single qutrit operations as a proof of concept.\\
\indent When compared to the existing qutrit CZ gates obtained using tunable cross-Kerr entanglement \cite{https://doi.org/10.48550/arxiv.2206.07216, https://doi.org/10.48550/arxiv.2206.11199}, our method has the potential to yield higher fidelities, as indicated by our results in Fig. \ref{fig:5}. Our method offers the advantage of giving full control using a single tunable parameter which is the flux of the coupler transmon. This also helps avoid sources of infidelity often associated with tunable circuit components since all the other transmons in the circuit can be made fixed frequency. For our chosen parameter set, the total execution time for realizing a CZ gate which would be 750 ns, not including time taken by single qutrit operations. However, we expect that future work on pulse optimization may lead to faster execution times. Our method for realizing the gates we refer to as iSWAP$_{01,10}$ and iSWAP$_{12,21}$ perform similarly to the iSWAP like gates explored in \cite{PhysRevA.108.032615}. However, our work shows an explicit CZ decomposition which is very useful in realization of the qutrit Clifford group. Our approach can also be extended to entangling of higher dimensional qudits by adding gates of form iSWAP$_{d-1\text{ }d,d\text{ }d-1}$. An important property of these gates is that they can be independently operated since these interactions are ideally limited within independent subspaces
\section*{Acknowledgments}
The authors acknowledge Vladyslov Los and Xi Dai for useful discussions. AL acknowledges funding support through an NSERC Discovery grant and an NSERC Alliance Quantum International grant and MS acknowledges the MITACS Globalink Scholarship.

\appendix

\section{Hamiltonian calculation from the circuit}\label{ap:capacit}
In this section we obtain the Hamiltonian for the circuit geometry represented in Fig. 1(a). Using the generalized coordinates $\phi_1$, $\phi_2$ and $\phi_c$, which are the phases across each of the junctions, and introducing for convenience the vector $\pmb{\phi}^T = \begin{bmatrix}
    \phi_1 &\phi_{\mathrm{c}} & \phi_2,
\end{bmatrix}$ we can define the Lagrangian
\begin{gather}
    L = \frac{1}{2}\pmb{\dot{\phi}}^T \mathbf{C}\pmb{\dot{\phi}} + \sum_{i=\mathrm{1,2}}E_{J_i}\cos{\phi_i} + E_{J_{\mathrm{c}}}|\cos(\Phi)|\cos{\phi_{\mathrm{c}}},
\end{gather}
with the capacitance matrix $\mathbf{C}$ given by
\begin{gather}
    \mathbf{C} = \begin{pmatrix}
        C_1 + C_{1\mathrm{c}} + C_{12} & -C_{1\mathrm{c}} & -C_{12}\\
        -C_{1\mathrm{c}} & C_{\mathrm{c}} + C_{1\mathrm{c}} + C_{2\mathrm{c}}  & -C_{2\mathrm{c}}\\
        -C_{12} & -C_{2\mathrm{c}} & C_2 + C_{2\mathrm{c}} + C_{12}
    \end{pmatrix}.
\end{gather}
The conjugate momenta $\mathbf{q} = \dfrac{\partial L}{\partial \dot{\pmb{\phi}}}$ are given by $\mathbf{q} = \mathbf{C}\pmb{\phi}$. The Hamiltonian, obtained using a Legendre transformation, is given by
\begin{equation}
\begin{aligned}
    H &= \sum_{i=1,2,\mathrm{c}}q_i\dot{\phi}_i - L = \frac{1}{2}\mathbf{q}^T\mathbf{C}^{-1}\mathbf{q}\\ 
    &- \sum_{i=\mathrm{1,2}}E_{J_i}\cos{\phi_i} - E_{J_{\mathrm{c}}}|\cos(\Phi)|\cos{\phi_{\mathrm{c}}}.
\end{aligned}
\end{equation}
The capacitance matrix is the same as the one used in Ref.~\cite{PhysRevApplied.10.054062}. Working in the limit of $C_{12}\ll C_{i\mathrm{c}}\ll C_i\sim C_{\mathrm{c}}$ where $i=1,2$, we obtain the following approximate inverse of the capacitance matrix:
\begin{equation}
    \mathbf{C}^{-1}\approx \begin{pmatrix}
        \frac{1}{C_1} & \frac{C_{1\mathrm{c}}}{C_1C_{\mathrm{c}}} & \frac{C_{12} + (C_{1\mathrm{c}}C_{2\mathrm{c}})/C_{\mathrm{c}}}{C_1C_2}\\
        \frac{C_{1\mathrm{c}}}{C_1C_{\mathrm{c}}} & \frac{1}{C_\mathrm{c}} & \frac{C_{2\mathrm{c}}}{C_2C_{\mathrm{c}}}\\
        \frac{C_{12} + (C_{1\mathrm{c}}C_{2\mathrm{c}})/C_{\mathrm{c}}}{C_1C_2} & \frac{C_{2\mathrm{c}}}{C_2C_{\mathrm{c}}} & \frac{1}{C_\mathrm{c}}
    \end{pmatrix}.
\end{equation}
We can see that the diagonal terms of this matrix simply correspond to the capacitive energy $E_{C_i} = e^2/2C_{i}$ for a given transmon with a capacitor of capacitance $C_i$. The Hamiltonian parameters used for simulation can be achieved by the following capacitor values.
\begin{table}[ht]
    \centering
    \begin{tabular}{|c|c|c|c|c|c|c|}
    \hline
      Capacitor & $C_1$ & $C_2$ & $C_{\mathrm{c}}$ & $C_{1\mathrm{c}}$ & $C_{1\mathrm{c}}$ & $C_{12}$\\
       \hline
       fF & 69.055 & 80.564 & 87.888 & 5.728 & 7.597 & 0.045\\
    \hline
    \end{tabular}
    \caption{Simulation parameters}
    \label{tab:2}
\end{table}

\section{The Schreiffer-Wolff frame}\label{ap:SWTframe}
The Schreiffer-Wolff frame is a reference frame where the Hamiltonian is block diagonalized over some subspace. In our case, we define it as a frame where interactions caused due to $\hat{H}_{\mathrm{m}}$ have been eliminated upto some order in $g_1/\Delta$ where $\Delta$ is the energy scale of $\hat{H}_0$ (the difference between the first excited state energy of the Hamiltonian to its ground state). The main approach we use for numerical methods is based on Ref.~\cite{PhysRevA.101.052308}. We first examine the more commonly used time-independent version of the Schreiffer-Wolff transformation, induced by a transformation $e^S$. The Hamiltonian in the new frame is
\begin{equation}\label{eq:SWTtime}
\begin{aligned}
    \hat{H}_\mathrm{SW} &= e^S\hat{H}_\mathrm{lab}e^{-S}\\ &= \hat{H}_\mathrm{lab} + [S,\hat{H}_\mathrm{lab}] + \frac{1}{2}[S,[S,\hat{H}_\mathrm{lab}]] + \dots .
\end{aligned}
\end{equation}
The ideal transformation would make $\hat{H}_\mathrm{SW}$ block-diagonal over Q1 and Q2 removing interactions to C. This can be calculated perturbatively by assuming $\hat{H}_{\mathrm{m}}$ to be much smaller in magnitude than $\hat{H}_0$. In this approach we take $S = \sum_{i=1}^{\infty}S^{(i)}$. If $\{\ket{i}\}$ is the basis set which diagonalizes $\hat{H}_0$ with eigenvalues $\{\lambda_i\}$, we can define the following first order SWT
\begin{equation}
    \langle{i|S^{(1)}|j}\rangle = \begin{cases}\dfrac{\langle{i|H_1|j}\rangle}{\lambda_i - \lambda_j} & \text{if and only if } i\neq j \text{ and } \lambda_i \neq \lambda_j\\0 & \text{otherwise}\end{cases} .\label{eq:A12}
\end{equation}

\indent In the time dependent case, discusssed in Ref.~\cite{PhysRevA.102.042605}, which is suitable given that the system Hamiltonian  $\hat{H}_\mathrm{lab}$ is time-dependent, the  effective Hamiltonian which evolves the state $\ket{\psi_\mathrm{SW}(t)} = e^{S(t)}\ket{\psi_\mathrm{lab}(t)}$ is given by
\begin{equation}
\begin{aligned}
    \hat{H}_\mathrm{SW}(t) &= e^{S(t)}(\hat{H}_\mathrm{lab}-i\partial_t)e^{-S(t)}\\ &= e^{S(t)}(\hat{H}_\mathrm{lab}+i(\partial_tS(t)))e^{-S(t)} .
\end{aligned}
\end{equation}
To obtaine a block diagonal form in this case, we need to solve a differential equation for $S(t)$ which has dependence on $\hat{H}_\mathrm{lab}(t)$. A complete description of all the terms up to higher orders can be found in Ref.~\cite{PhysRevA.102.042605}. The expansion of $\hat{H}_\mathrm{SW}(t)$ is as follows (without terms involving orders of $S^{(2)}$ and higher):
\begin{equation}
\begin{aligned}
    \hat{H}_\mathrm{SW}(t) &= \hat{H}_{0}(t) + \left(\hat{H}_{\mathrm{m}} + [S^{(1)},\hat{H}_0] + i\dot{S}^{(1)}\right) \\&+ \hat{H}_{\mathrm{d}} + \frac{1}{2}[S^{(1)},\hat{H}_{\mathrm{m}}]+\dots .
\end{aligned}
\end{equation}
The following equation needs to be solved for $S^{(1)}(t)$:
\begin{equation}\label{eq:SWTder}
    \dot{S}^{(1)} = i(\hat{H}_{\mathrm{m}} + [S^{(1)},\hat{H}_0]) .
\end{equation}
For simplicity we represent the Hamiltonian $\hat{H}_\mathrm{SW}(t) = \hat{H}_\mathrm{SW,0}(t) + \hat{H}_\mathrm{SW,c}(t)$. Here $\hat{H}_{\mathrm{SW,0}}(t)$ represents the diagonal part of $\hat{H}_{\mathrm{SW}}(t)$ and $\hat{H}_\mathrm{SW,c}(t)$ represents the off-diagonal part when expressed in the basis which diagonalizes $\hat{H}_0(t=0)$.
 
\section{The rotating frame}\label{ap:rotframe}
As defined in equation \eqref{eq:finalTransform}, we make use of 
a rotating frame for the computation. The state in this frame is defined as $\ket{\psi_\mathrm{rot}(t)} = U_\mathrm{rot}(t)\ket{\psi_\mathrm{lab}(t)}$. To capture the evolution in this frame we first write the evolution in the lab frame:
\begin{equation}
    i\hbar\partial_t\ket{\psi_\mathrm{lab}(t)} = \hat{H}_\mathrm{lab}\ket{\psi_\mathrm{lab}(t)}.
\end{equation}
The evolution in the transformed frame defined by transformation in equation \eqref{eq:finalTransform} is given as follows
\begin{gather}
    i\hbar\partial_t\ket{\psi_\mathrm{rot}(t)} = \hat{H}_\mathrm{rot}(t)\ket{\psi_\mathrm{rot}(t)}\\
    \hat{H}_\mathrm{rot}(t) = U_\mathrm{rot}(t)H_\mathrm{lab}(t)U^\dagger_\mathrm{rot}(t) - i\hbar U_\mathrm{rot}(t)\partial_tU_\mathrm{rot}^\dagger(t)
\end{gather}
For shorthand notation, we will refer $\int_0^t \hat{H}_\mathrm{SW,0}(t')dt'/\hbar$ as $\mathcal{H}(t)$. We have
\begin{equation}
\begin{aligned}
    i\hbar U_\mathrm{rot}(t)\partial_tU_\mathrm{rot}^\dagger(t) &= i\hbar e^{i\mathcal{H}(t)}e^{S}(\partial_t S)e^{-S}e^{-i\mathcal{H}(t)}\\ &\quad+  i\hbar e^{i\mathcal{H}(t)}e^{S}e^{-S}(-i\partial_t \mathcal{H}(t))e^{-i\mathcal{H}(t)}
\end{aligned}
\end{equation}
which can be simplified to obtain
\begin{equation}
\begin{aligned}
    i\hbar U_\mathrm{rot}(t)\partial_tU_\mathrm{rot}^\dagger(t) &= i\hbar e^{i\mathcal{H}(t)}e^{S}(\partial_t S)e^{-S}e^{-i\mathcal{H}(t)}\\ &\quad+  e^{i\mathcal{H}(t)}\hat{H}_\mathrm{SW,0}(t)e^{-i\mathcal{H}(t)}.
\end{aligned}
\end{equation}
From equation \eqref{eq:SWTtime}, we have $U_\mathrm{rot}(t)H_\mathrm{lab}(t)U_\mathrm{rot}^\dagger(t) = e^{i\mathcal{H}(t)}(\hat{H}_\mathrm{SW}(t) + e^{S(t)}(\partial_t S(t))e^{-S(t)})e^{-i\mathcal{H}(t)}$. This gives us the following expression for $\hat{H}_\mathrm{rot}(t)$:
\begin{equation}
    \hat{H}_\mathrm{rot}(t) = e^{i\mathcal{H}(t)}\hat{H}_\mathrm{SW,c}(t)e^{-i\mathcal{H}(t)}
\end{equation}

\section{Equivalent Hamiltonian for AC drive}\label{ap:Eqrotframe}
Here we examine a flux drive from $t=0$ to $t=T$ of the form
\begin{equation}
    \Phi(t) = \Phi_0 + e(t)\delta\cos(\omega t).
\end{equation}
Here $e(t)$ is some envelope for which $e(0) = e(T) = 0$. For sake of simplification we will assume at first thatthat $e(t) = \Theta(t)\Theta(T-t)$ where $\Theta(x)$ is the Heaviside step function. Given that $\hat{H}_{\mathrm{SW,c}}(t)$ has a dependence on flux and hence on time, we can perform a Taylor expansion to obtain
\begin{equation}
    \begin{aligned}
        \hat{H}_{\mathrm{SW,c}}(\Phi) &= \hat{H}_{\mathrm{SW,c}}(\Phi_0) + (\Phi - \Phi_0)\frac{\partial\hat{H}_{\mathrm{SW,c}}(\Phi_0)}{\partial\Phi} \\
    &\quad+ \frac{(\Phi - \Phi_0)^2}{2}\frac{\partial^2\hat{H}_{\mathrm{SW,c}}(\Phi_0)}{\partial\Phi^2} + \dots ,
    \end{aligned}
\end{equation}
which can also be written as
\begin{equation}\label{eq:taylorHSWc}
    \begin{aligned}
      \hat{H}_{\mathrm{SW,c}}(\Phi) &= \hat{H}_{\mathrm{SW,c}}(\Phi_0) + \frac{\delta^2}{4}\frac{\partial^2\hat{H}_{\mathrm{SW,c}}(\Phi_0)}{\partial\Phi^2} \\
    &\quad+ \delta\cos{\omega t}\left(\frac{\partial\hat{H}_{\mathrm{SW,c}}(\Phi_0)}{\partial\Phi}\right) \\
    &\quad+ \frac{\delta^2\cos{(2\omega t)}}{4}\left(\frac{\partial^2\hat{H}_{\mathrm{SW,c}}(\Phi_0)}{\partial\Phi^2}\right) + \mathcal{O}(\delta^3).
    \end{aligned}
\end{equation}
The Hamiltonian $\hat{H}_{\mathrm{SW,0}}$ can be expanded in a similar way. We assume that we operate in the basis set $\{\ket{i}\}$ which diagonalizes $\hat{H}_0(\Phi_0)$. In this case the matrix elements of $\hat{H}_\mathrm{rot}(t)$ can be expressed as
\begin{equation}
    \langle i|\hat{H}_\mathrm{rot}(t)|j\rangle = \sum_{k,l}\langle i|e^{i\mathcal{H}(t)}|k\rangle\langle k|\hat{H}_\mathrm{SW,c}(t)|l\rangle\langle l|e^{-i\mathcal{H}(t)}|j\rangle .
\end{equation}
Note that $e^{i\mathcal{H}(t)}$ is diagonal in the $\{\ket{i}\}$ basis. This gives us the following result:
\begin{equation}
    \langle i|\hat{H}_\mathrm{rot}(t)|j\rangle = \langle i|\hat{H}_\mathrm{SW,c}(t)|j\rangle\exp(i(\mathcal{H}(t)_{i,i} - \mathcal{H}(t)_{j,j}))
\end{equation}
Here $\mathcal{H}(t)_{i,i} = \langle i|\mathcal{H}(t)|i\rangle$. We can see from the form in equation \eqref{eq:taylorHSWc} that 
\begin{gather}
    \mathcal{H}(t)_{i,i} = \omega_it + \delta\left(\frac{\partial\hat{H}_{\mathrm{SW,0}}(\Phi_0)}{\partial\Phi}\right)\int_0^t\cos{\omega t'}dt' + \mathcal{O}(\delta^2)\\
    \omega_i = \hat{H}_{\mathrm{SW,0}}(\Phi_0) + \frac{\delta^2}{4}\frac{\partial^2\hat{H}_{\mathrm{SW,0}}(\Phi_0)}{\partial\Phi^2}+\mathcal{O}(\delta^3)
\end{gather}
Here we can clearly see that the integrals of the rotating terms can be very easily bounded by a constant value. On taking the approximation that $\mathcal{H}(t)_{i,i} \approx \omega_it$, we  note that the following holds with $\Delta_{i,j} = \omega_i - \omega_j$
\begin{equation}
\begin{aligned}
    \langle i|\hat{H}_\mathrm{rot}(t)|j\rangle &\approx \langle i|\hat{H}_\mathrm{SW,c}(t)|j\rangle e^{i\Delta_{i,j} t}\\ &= \frac{(e^{(\Delta_{i,j} - \omega)t}  + e^{(\Delta_{i,j} + \omega)t})}{2}\frac{\partial\langle i|\hat{H}_{\mathrm{SW,c}}(\Phi_0)|j\rangle}{\partial\Phi} \\&\quad + \mathcal{O}(\delta^3).
\end{aligned}
\end{equation}
The other terms in the above equation will be rotating terms however we can note that in the condition where $\Delta_{ij} = \pm\omega$, there will be a non-rotating term which can result in an interaction between $\ket{i}$ and $\ket{j}$. Using the rotating wave approximation, we can note that the equivalent Hamiltonian in the rotating frame is
\begin{equation}
    \hat{H}_{\mathrm{equiv}} = \left(\frac{\partial\langle i|\hat{H}_{\mathrm{SW,c}}(\Phi_0)|j\rangle}{\partial\Phi}\right)\frac{(\ket{i}\bra{j} + \ket{j}\bra{i})}{2} .
\end{equation}
\section{Qutrit CZ decompostion}
The following are the circuit parameters for an exact qutrit CZ decomposition with angles rounded to 4 decimal places as a fraction of $\pi$.\\
-----------------------------------------------\\
--------------Initial Rotation--------------\\
On subspace $\ket{0}$, $\ket{1}$ of qudit 1\\
\indent RX(0.3584$\pi$)\\
\indent RZ(-0.0$\pi$)\\
\indent RX(0.6416$\pi$)\\
On subspace $\ket{0}$, $\ket{1}$ of qudit 2\\
\indent RX(-0.4014$\pi$)\\
\indent RZ(-1.0$\pi$)\\
\indent RX(-0.4014$\pi$)\\
On subspace $\ket{0}$, $\ket{2}$ of qudit 1\\
\indent RX(0.5664$\pi$)\\
\indent RZ(0.7678$\pi$)\\
\indent RX(0.088$\pi$)\\
On subspace $\ket{0}$, $\ket{2}$ of qudit 2\\
\indent RX(-0.4999$\pi$)\\
\indent RZ(-0.4992$\pi$)\\
\indent RX(-0.0375$\pi$)\\
On subspace $\ket{1}$, $\ket{2}$ of qudit 1\\
\indent RX(1.0$\pi$)\\
\indent RZ(-0.4539$\pi$)\\
\indent RX(0.0$\pi$)\\
On subspace $\ket{1}$, $\ket{2}$ of qudit 2\\
\indent RX(0.7868$\pi$)\\
\indent RZ(0.9999$\pi$)\\
\indent RX(-0.2132$\pi$)\\
\\
------------------Layer 1------------------\\
Apply iSWAP1(-0.6667$\pi$)\\
On subspace $\ket{0}$, $\ket{1}$ of qudit 1\\
\indent RX(0.196$\pi$)\\
\indent RZ(-0.6794$\pi$)\\
\indent RX(1.1255$\pi$)\\
On subspace $\ket{0}$, $\ket{1}$ of qudit 2\\
\indent RX(0.7722$\pi$)\\
\indent RZ(-0.0264$\pi$)\\
\indent RX(-0.232$\pi$)\\
On subspace $\ket{0}$, $\ket{2}$ of qudit 1\\
\indent RX(-0.5699$\pi$)\\
\indent RZ(0.0$\pi$)\\
\indent RX(-0.4301$\pi$)\\
On subspace $\ket{0}$, $\ket{2}$ of qudit 2\\
\indent RX(0.5671$\pi$)\\
\indent RZ(-1.0$\pi$)\\
\indent RX(-0.4329$\pi$)\\
On subspace $\ket{1}$, $\ket{2}$ of qudit 1\\
\indent RX(-0.1632$\pi$)\\
\indent RZ(-0.5265$\pi$)\\
\indent RX(-0.0092$\pi$)\\
On subspace $\ket{1}$, $\ket{2}$ of qudit 2\\
\indent RX(0.555$\pi$)\\
\indent RZ(-0.7193$\pi$)\\
\indent RX(0.0225$\pi$)\\
\\
------------------Layer 2------------------\\
Apply iSWAP2(0.6667$\pi$)\\
On subspace $\ket{0}$, $\ket{1}$ of qudit 1\\
\indent RX(-0.6542$\pi$)\\
\indent RZ(-1.0$\pi$)\\
\indent RX(-0.6542$\pi$)\\
On subspace $\ket{0}$, $\ket{1}$ of qudit 2\\
\indent RX(0.0$\pi$)\\
\indent RZ(0.3333$\pi$)\\
\indent RX(-1.0001$\pi$)\\
On subspace $\ket{0}$, $\ket{2}$ of qudit 1\\
\indent RX(-0.0001$\pi$)\\
\indent RZ(0.8795$\pi$)\\
\indent RX(-0.0001$\pi$)\\
On subspace $\ket{0}$, $\ket{2}$ of qudit 2\\
\indent RX(-0.1243$\pi$)\\
\indent RZ(-1.1$\pi$)\\
\indent RX(-0.6188$\pi$)\\
On subspace $\ket{1}$, $\ket{2}$ of qudit 1\\
\indent RX(1.1312$\pi$)\\
\indent RZ(0.6862$\pi$)\\
\indent RX(0.5754$\pi$)\\
On subspace $\ket{1}$, $\ket{2}$ of qudit 2\\
\indent RX(0.0562$\pi$)\\
\indent RZ(1.0$\pi$)\\
\indent RX(0.0562$\pi$)\\

\end{document}